\documentclass[superscriptaddress, reprint]{revtex4-2} 
\usepackage{graphicx}
\usepackage{amssymb}
\usepackage{epstopdf}
\usepackage{amsmath,dsfont, commath}
\usepackage{bm}
\usepackage{braket}
\usepackage{changes}
\usepackage{hyperref}

\newcommand{\be}{\begin{equation}}
	\newcommand{\ee}{\end{equation}}
\newcommand{\bea}{\begin{eqnarray}}
	\newcommand{\eea}{\end{eqnarray}}

\newcommand{\ba}{\begin{array}}
	\newcommand{\ea}{\end{array}}

\newcommand{\se}{Schr\"{o}dinger equation}
\newcommand{\bl}{\begin{flalign}}
	\newcommand{\enl}{\end{flalign}}

\newcommand{\mc}[1]{\mathcal{#1}}

\newcommand{\tdse}{time-dependent Schr\"{o}dinger equation}

\newcommand{\eq}[1]{Eq. \eqref{#1}}

\newcommand{\fig}[1]{Fig. \ref{#1}}

\usepackage{siunitx}
\usepackage[capitalize]{cleveref}

\newcommand{\proj}[1]{\ket{#1}\bra{#1}}

\renewcommand{\bf}[1]{\mathbf{#1}}
\newcommand{\grad}{\nabla}

\begin{document}
	
	\title{Path-ordered linked product approximation to the global electronic overlap matrix}
\author{Yujuan Xie}
\author{Bing Gu}
\email{gubing@westlake.edu.cn}
\affiliation{Institute of Natural Sciences, Westlake Institute for Advanced Study, Hangzhou, Zhejiang 310024, China}
\affiliation{Department of Chemistry, School of Science, Westlake University, Hangzhou, Zhejiang 310030, China}

\begin{abstract}

The global many-electron wave function overlap matrix accounts for all effects beyond the Born-Oppenheimer approximation in the discrete variable local diabatic representation, a numerically exact framework for modeling nonadiabatic conical intersection wave packet dynamics. Nevertheless, calculating the electronic overlap matrix from electronic structure is computationally expensive. Here, we introduce an approximation for constructing the electronic overlap matrix between any two long-range geometries by the product of nearest-neighbor overlap matrices along a path connecting these two geometries. This approximation significantly reduces the computational effort by only requiring electronic structure calculations for the nearest-neighbor overlap matrices. The accuracy of this approximation is demonstrated through an exact simulation of a proton-coupled electron transfer model. Our results show that although the approximate overlap matrix can exhibit noticeable differences from the exact ones, the conical intersection dynamics is in almost exact agreement with those from the exact overlap matrix.
\end{abstract}

\maketitle

\section{Introduction}

Conical intersections (CIs) are critical points where two or more adiabatic potential energy surfaces of a molecule intersect, leading to nonadiabatic transitions between adiabatic electronic states and geometric phase effects. CIs play the role of a ``transition state'' in a wide range of photochemical, photobiological, and photophysical processes, such as nonradiative electronic relaxation, energy transfer, singlet fission, intersystem crossing, and chemical reactions \cite{domcke2011, larson2020}.
Accurate modeling of photoexcited nuclear wave packet dynamics traversing through conical intersections is important for understanding and predicting the behavior of molecules upon photoexcitation and in time-resolved spectroscopy \cite{worner2011, kowalewski2017, gu2020b, gu2020c, yang2020b, xie2016}. Directly simulating CI dynamics with adiabatic electronic states face a fundamental challenge that the adiabatic electronic states cannot be made globally smooth, causing singularities in the first- and second-order derivative couplings at CIs \cite{han2023, zhu2024}. Hypothetically, this problem can be addressed by a geometry-dependent unitary transformation of the adiabatic electronic states to diabatic states, whereby derivative couplings vanish.
However, such exact diabatization does not exist for a finite number of electronic states, so approximate quasi-diabatizations are often used instead to transform away the singular component of the derivative couplings \cite{aleotti2021b, shu2022}, neglecting the residual derivative couplings.

The discrete variable local diabatic representation (LDR) is a numerically exact yet divergence-free framework,  recently proposed for simulating nonadiabatic conical intersection dynamics \cite{gu2023b, gu2024a, zhu2024}. In it, the Global many-Electron wave function Overlap Matrix (GEOM) —the overlap between adiabatic electronic states  at different molecular configurations—accounts for all  effects beyond Born-Oppenheimer, including nonradiative electronic relaxation, geometric phase effects, and diagonal Born-Oppenheimer corrections, thus paving the way for exact molecular quantum dynamics. 

Computing the global electronic overlap matrix from quantum chemistry is currently the bottleneck of the LDR method, even for a few nuclear degrees of freedom, as it requires the overlap matrix between any two nuclear configurations. With a direct-product discrete variable representation, the nuclear basis set (and grid points) for a $d$-dimensional conical intersection dynamics simulation scales as $\mc{O}(n^{d})$, where $n$ is the typical number of grid points for each degree of freedom. The required computational cost of the overlap matrix then scales as $\mc{O}(n^{2d})$.  

Here we introduce a linked product approximation using path-ordered products of nearest-neighbor electronic overlap matrix (referred to as a link) to construct the global overlap matrix. With this approximation, explicit electronic structure calculations are only required for the overlap matrix between a configuration $\bf R$ and its $d$ nearest-neighbors. Hence, the computational cost scaling reduces to $\mc{O}\del{d n^d}$. The remaining matrix elements between non-nearest-neighbor configurations are obtained through a path-ordered product of links along a connecting path.

The accuracy of the linked product approximation is demonstrated by a numerically exact ``ab initio" simulation of a two-dimensional Shin-Metiu model, a photo-induced proton-coupled electron transfer model \cite{shin1995, shin2000}. It is  shown that the quantum dynamics results obtained using the approximate overlap matrix are in almost exact agreement with those obtained using the exact overlap matrix. This demonstrates that the linked product approximation reduces computational cost yet without sacrificing the accuracy of the LDR method.

This paper is organized as follows. In \cref{sec:method}, we briefly describe the LDR method and show the critical roles played by the many-electron overlap matrix in CI dynamics. We then introduce the linked product approximation. An efficient Implementation and a numerical demonstration for a conical intersection model is shown in \cref{sec:resutls&discussions}.  \cref{sec:summary} concludes.

Atomic units $\hbar = e = m_\text{e} = \si{\bohr} = 1$ are used throughout.

\section{Theory} \label{sec:method}
\subsection{Electronic overlap matrix in the Local diabatic representation}
We briefly present how the global electronic overlap enters into the nonadiabatic CI dynamics in the LDR method. The LDR employs a discrete variable representation (DVR) for the nuclear coordinate operators \cite{light2000, littlejohn2002}. That is, we use DVR basis sets to describe the nuclear motion. Any DVR basis set can be used directly in LDR; the optimal choice depends on the specific problem and boundary conditions. The DVR basis set has been widely used in computing vibrational eigenstates and wave packet dynamics on a single adiabatic potential energy surface and is remarkably accurate with exponential convergence \cite{littlejohn2002}.

In contrast to the widely used Born-Huang expansion for the total molecular wavefunction, LDR employs an ansatz that removes the explicit nuclear dependence of the electronic states,
\be
\Psi(\bf r, \bf R, t) = \sum_{\bf n} \sum_\alpha C_{\bf n \alpha}(t) \phi_\alpha(\bf r; \bf R_{n}) {\chi_{\bf n}}(\bf R) 
\label{eq:ansatz}
\ee   
where $\bf R_n$ is a nuclear configuration, labeled by a multi-index $\bf n = \set{n_1, n_2, \dots, n_d}$ and predetermined by the DVR of the nuclear coordinates, and $\alpha$ runs over the adiabatic electronic states relevant to the chemical process. In \cref{eq:ansatz}, $\chi_n(\bf R)$ is the DVR set.  
The total molecular wave function is expanded by the vibronic basis set $ \{ \phi_\alpha(\bf r; \bf R_{n}) {\chi_n}(\bf R) \}$ with expansion coefficients $C_{n\alpha}(t)$.

LDR inherits the advantages of a DVR basis set that both the nuclear kinetic energy operators and potential energy matrix elements can be easily computed. The kinetic energy operator matrix elements can be computed analytically, and the potential energy operator matrix elements are calculated using a diagonal approximation, avoiding high-dimensional quadrature.

For a $d$-dimensional direct product DVR, the nuclear geometries are simply generated by the direct product for each coordinate 
\be
\ket{\bf R_{\bf n}} = \ket{R_{n_1}^1} \otimes \ket{R_{n_2}^2} \cdots \otimes \ket{R_{n_d}^d}.
\ee 
The adiabatic electronic states $\ket{\phi_{\alpha}(\bf{R_n})}$ are eigenstates of the electronic Hamiltonian  
\be 
 \hat{H}_\text{e}(\bf R) = \hat{T}_\text{e} + \hat{V}_\text{ee} + \hat{V}_\text{eN}(\bf{r}; \bf{R})  + \hat{V}_\text{NN}(\bf R) ,
 \ee
  consisting of electronic kinetic energy $\hat{T}_\text{e}$, electron-nuclear interaction $\hat{V}_\text{en}$, electron-electron Coulomb repulsion $\hat{V}_\text{ee}$ and nuclear repulsion energy $\hat{V}_\text{nn}$, 
\be  \hat{H}_\text{e}(\bf r; \bf R_n)\phi_\alpha(\bf r; \bf R_n) = V_\alpha(\bf R_n) \phi_\alpha(\bf r; \bf R_n) 
.\ee

The propagation of the molecular \tdse\ given an initial state $\ket{\Psi_0}$, i.e., $\ket{\Psi(t)} = e^{-i \hat{H} t} \ket{\Psi_0}$, is performed by applying the second-order Strang splitting to the short-time {propagator} \cite{gu2024}.

\be
e^{-i \hat{H} \Delta t} =  e^{- i \hat{H}_\text{e}(\bf R) \Delta t/2}   e^{- i \hat{T}_\text{N} \Delta t} e^{- i \hat{H}_\text{e}(\bf R) \Delta t/2} + \mc{O}(\Delta t^3)
\ee
The wave function propagation with electronic Hamiltonian $\hat{H}_e$ yields 
\begin{equation}
 e^{-i \hat{H}_\text{e}(\bf R) \Delta t}  \ket{\bf n \alpha}  \approx 
	 e^{-i V_{n\alpha} \Delta t } \ket{\bf n\alpha}
\end{equation}
where $V_{\bf n \alpha} = V_{\alpha}(\bf R_{\bf n)}$ is the electronic energy and $\ket{\bf n \alpha} \equiv \ket{\phi_\alpha(\bf R_{\bf n})} \otimes \ket{\chi_{\bf n}}$ is a shorthand notation. 
As $\hat{T}_\text{N}$ only operates on  nuclear basis functions, the nuclear kinetic energy propagator is 
\begin{align}
	 &\braket{ \bf{m} \beta| e^{- i \hat{T}_\text{N}\Delta t} | \bf{n} \alpha} 
	= {A}_{\bf m, \bf n}^{\beta \alpha} \braket{ \bf{m} | e^{-i \hat{T}_\text{N} \Delta t}|  \bf{n}}_{\bf R}
\end{align}
where the GEOM is defined by the overlap between two adiabatic states of reference geometries $\bf{R_m}$ and  $\bf{R_n}$
\begin{equation}
	{A}_{\bf m \bf n}^{\beta \alpha} = \braket{\phi_\beta (\bf{R_m}) | \phi_\alpha (\bf{R_n})}_\bf{r}
\end{equation}
${\langle \cdots \rangle}_\bf{R}$ and ${\langle \cdots \rangle}_\bf{r}$ denote the integration over nuclear and electronic degrees of freedom respectively.
The $\braket{\bf{m} | e^{-i \hat{T}_\text{N} \Delta t}| \bf{n}}_{\bf R}$ represents the matrix element of exponential kinetic energy operator, which can be easily calculated in the DVR basis sets.

The definition of the electronic overlap matrix restricts its matrix elements to the range of $[-1, 1]$. Since the GEOM capturing all effects beyond Born-Oppenheimer is singularity-free, the LDR method is numerically robust even in the presence of electronic degeneracies including conical intersections. The adiabatic electronic energy $V_{\bf n \alpha}$ and the electronic overlap matrix can be calculated through standard electronic structure methods, including density functional theory and post-Hartree-Fock methods. Usually, multi-configurational methods are required to describe CIs \cite{lu2022, yang2016}.

As the potential energy matrix is diagonal and the kinetic energy operator can be factorized, the most computationally expensive part is the overlap matrix. Even for a few degrees of freedom, the global electronic overlap can be computationally intractable.

\subsection{Building the global overlap matrix from links}

The challenge in approximating the overlap matrix is that it carries a random sign of $\pm 1$ originating from the electronic structure computations. This reflects the $\mathbb{Z}_2$ gauge freedom for real-valued electronic wavefunctions.  The phase information, despite being random, cannot be removed as it is critical to account for the geometric phase effects. 
Any loop integral in the configuration space has to yield the correct geometric phase
\be 
\text{Tr}_{\bf r}\cbr{ \proj{\gamma_1^\alpha} \proj{\gamma_2^\alpha}\cdots \proj{\gamma_N^\alpha}   }  = (-1)^{\nu_\gamma} ,
\ee 
where $\nu_\gamma$ is the winding number of the loop $\gamma$, i.e., how many times it encircles the CIs.
Therefore, the GEOM a highly irregular function that cannot be fitted by standard fitting models. 

To overcome this challenge, we introduce a semi-local approximation to calculate the GEOM that is globally phase consistent, i.e.,  This approximation, in essence, approximates the long-range electronic overlap matrix using short-range ones. The phase information in the short-range overlap matrices suffices to ensure global phase consistency. Specifically, we first compute the overlap matrix between nearest-neighbors $\bf L_{\bf n, \pm j} \equiv \bf A_{\bf{n}, \bf{n} \pm \bf e_j}, j = 1, 2, \dots, d$ using electronic structure methods. For each pair of configurations, $\bf A_{\bf n \bf m} $ is an $ M \times M$ matrix, where $M$ is the number of electronic states. 

For configurations labeled by $\bf n$ and $\bf m$ that are not nearest-neighbors, we first find a path $\gamma$ that connects these two configurations and approximate the corresponding overlap matrix by a product of links along this path

\be 
\begin{split}
	\bf A_{\bf m \bf n}  
	\approx & \mc{P}_\gamma \prod_{k=0}^{L-1} \bf A_{\gamma_k, \gamma_{k+1}}
 \end{split}
 \label{eq:lpa}
\ee 
where $\mc{P}_\gamma$ is the path-ordering operator that places the matrix appearing earlier in the path to the right of later ones, $ L =  \norm{\bf n - \bf m}_1$ is the path length,  $\gamma_0 = \bf n $ and $ \gamma_{L} = \bf m$. 

\begin{figure}[htbp]
	\centering
	\includegraphics[width=0.4\textwidth]{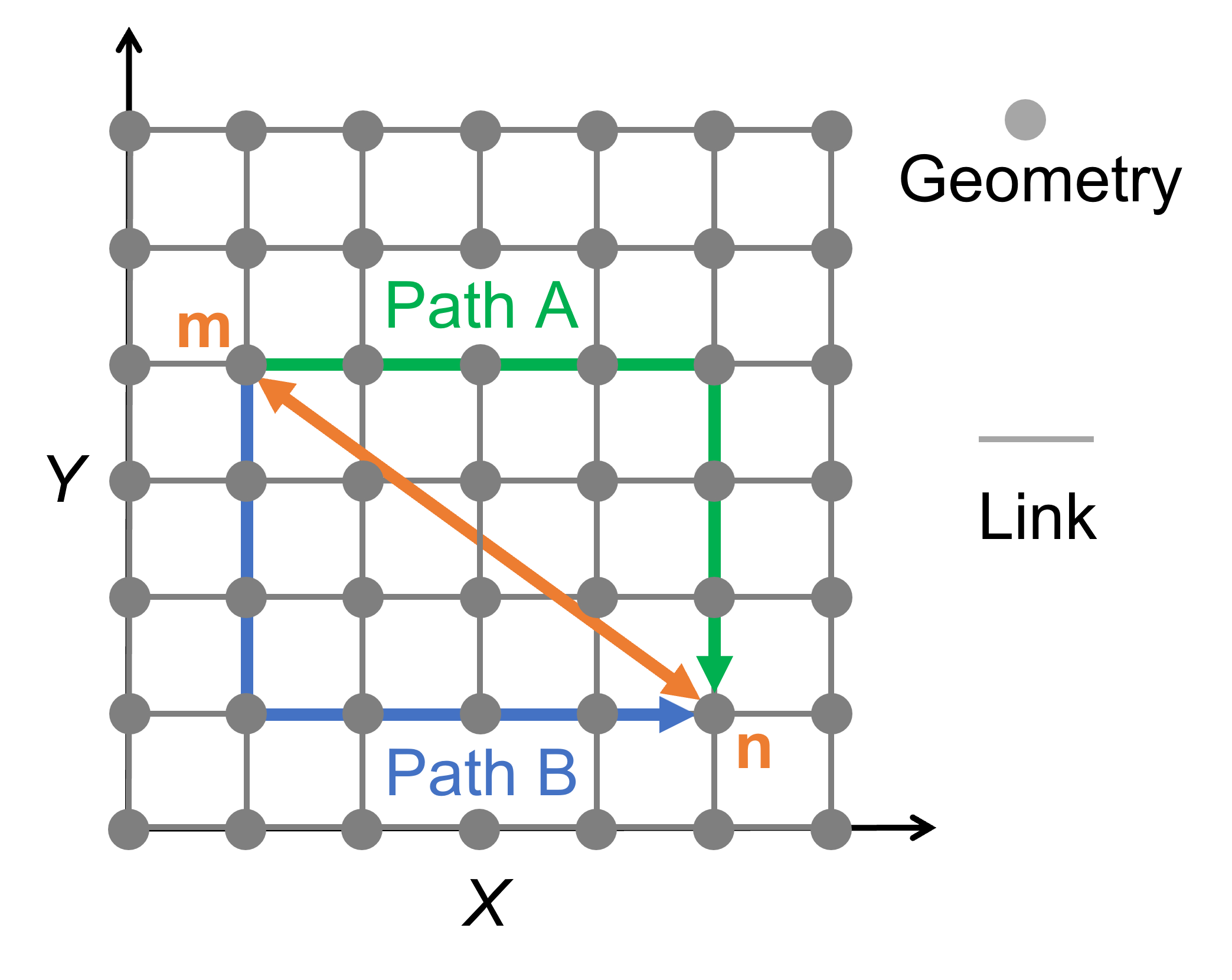}
	\caption{Schematic of calculating electronic overlap matrix between geometry $\bf m$ and geometry $\bf n$ by the linked product approximation through two pathways.}
	\label{fig:schematic}
\end{figure}

A schematic of this approximation is depicted in \fig{fig:schematic} for a two-dimensional configuration space.  Each node represents a configuration, and each ``bond'' is a link, i.e. the electronic overlap matrix between two nearest configurations. Two possible pathways that connect two long-range configurations are shown. 

This approximation can be understood as follows. Defining the electronic projection operators $\hat{P}_{\bf n} = \sum_\alpha \proj{\Phi_\alpha(\bf R_n)}$ at each nuclear configuration and its complement $\hat{Q}_{\bf n} = \bf 1 - \hat{P}_{\bf n}$, with $\bf 1$ as the electronic identity, and inserting the electronic identity for each node (configuration) along the path yields

\be 
\begin{split}
	\bf A_{\bf m \bf n} =& \Braket{\bm \Phi_{\bf m} | \prod_{k=1}^L \del{\hat{P}(\gamma_k) + \hat{Q}(\gamma_k)} | \bm \Phi_{\bf n}}_{\bf r}	\\ 
		\approx & \bf A_{\bf m \gamma_{M}}\cdots \bf A_{\gamma_{k+1}, \gamma_k}\cdots  \\ 
\end{split}
\label{eq:app}
\ee 
Neglecting the terms associated with the complementary operators $\hat{Q}$ leads to the linked product approximation in \eq{eq:lpa}. In the complete basis set limit, \cref{eq:lpa} is exact. 

For example, for a two-dimensional configuration space, to compute the electronic overlap matrix between configurations labeled by  (1,3) and (2,5), the path can be 
\be 
\gamma: (1,3) \rightarrow (2,3) \rightarrow (2,4) \rightarrow (2,5)
\ee 

The linked product approximation is path-dependent. There are many different paths that can connect two configurations. \eq{eq:app} implies that the shortest path should provide the best approximation. Nevertheless, this does not completely determine the path, as there are still many paths with the same shortest distance $L$, each providing an approximation at the same level of accuracy, as illustrated in \fig{fig:schematic}.
In this work, we selected two different paths, named Path A and Path B, to obtain two approximate overlap matrices, and then conducted quantum dynamics analysis based on these matrices.

Path A, connecting the configuration $\bf m$ to the configuration $\bf n$ in $d$-dimensions, is given by the sequence 
\be 
\begin{split}
\gamma_A: \bf m & \rightarrow  \bf (n_1, m_2, \cdots, m_d) \\
& \rightarrow (n_1, n_2, m_3, \dots, m_d) \\ 
& \rightarrow \cdots \rightarrow \bf n 
\end{split}
\ee 
In turn, path B corresponds to 
\be 
\begin{split}
\gamma_B: \bf m & \rightarrow  \bf (m_1, m_2, \cdots, n_d) \\
& \rightarrow (m_1, \dots, n_{d-1}, n_d) \\ 
& \rightarrow \cdots \rightarrow \bf n 
\end{split}
\ee

\subsection{Implementing the global overlap matrix by nearest-neighbors}
We use an recursive relation to implement the approximation for a $d$-dimensional system. The GEOM $\bf{A}$ can be obtained by 


\be
\mathbf{A}_{(d)} =\left( \sum_{k=1}^{N_d -1} \left( \mathbf{A}_{(d-1)} \left( \mathbf{L}_d \right) ^k + \left( \mathbf{L}_d \right) ^k \right) + \text{H. c.} \right) + \mathbf{A}_{(d-1)} + \bf I
\ee
where $N_d$ is the number of grid points in the $d$-th dimension. $\mathbf{A}_{(d-1)}$ is the overlap matrix for the $(d-1)$-dimensional system. $\bf{L}_d = \bf A_{\bf{n}, \bf{n} + \bf e_d}$ is the link along the positive direction of the $d$-th coordinate axis, i.e. the nearest-neighbor overlap matrix ,  
 $\bf I$ is a unit matrix, 
 and $\bf A_{(0)} = \bf 0$.

\begin{figure}[htbp]
	\centering
	\includegraphics[width=0.48\textwidth]{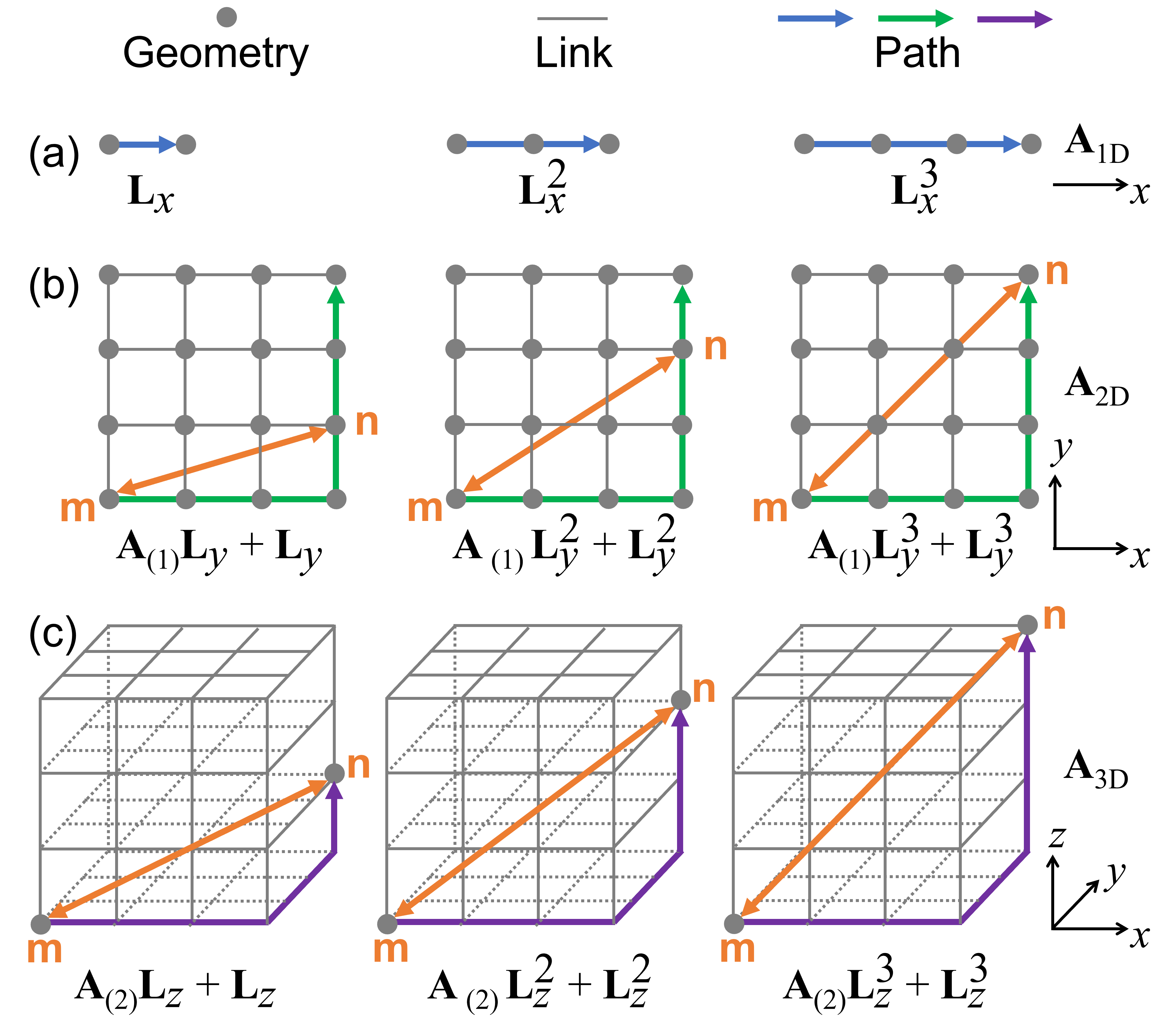}
	\caption{Schematic of calculating the global electronic overlap matrix for a three-dimensional system using the linked product approximation. (a), (b) and (c) represent the paths for $\bf A_{(1)}$, $\bf A_{(2)}$ and $\bf A_{(3)}$, respectively.} 
\label{implementation}
\end{figure}

Here we take a three-dimensional system as an example to illustrate the specific steps for computing the global overlap matrix:

(1) Calculate the overlap matrix between configurations that only differ in the $x$ axis (i.e., the $y$ and $z$ coordinates are the same, see \fig{implementation}a) using $\bf{L}_x$
\be
\bf A_{(1)} = \sum_{k=1}^{N_x-1} \left( \bf{L}_x \right)^k + \text{H. c.}
\ee

(2) Calculate the overlap matrix between configurations within the $x,y$ plane using $\bf{L}y$ and $\bf A_{(1)}$, where the $z$-coordinates of the configurations remain the same (see \fig{implementation}b)
\be
\bf A_{(2)} = \left( \sum_{k=1}^{N_y-1} \left( \bf A_{(1)} \left( \bf{L}_y \right)^k  + \left( \bf{L}_y \right)^k \right) + \text{H. c.} \right) + \bf A_{(1)}
\ee

(3) Finally, compute the overlap matrix for configurations in the three-dimensional space using $\bf{L}z$ and $\bf A_{(2)}$ (see \fig{implementation}c)
\be
\bf A_{(3)} = \left( \sum_{k=1}^{N_z-1} \left( \bf A_{(2)} \left( \bf{L}_z \right)^k  + \left( \bf{L}_z \right)^k \right) + \text{H. c.} \right) + \bf A_{(2)} + \bf I
\ee



%
%
%
%

\section{Model and Computations} \label{sec:resutls&discussions}

\begin{figure}[htbp]
	\centering
	\includegraphics[width=0.48\textwidth]{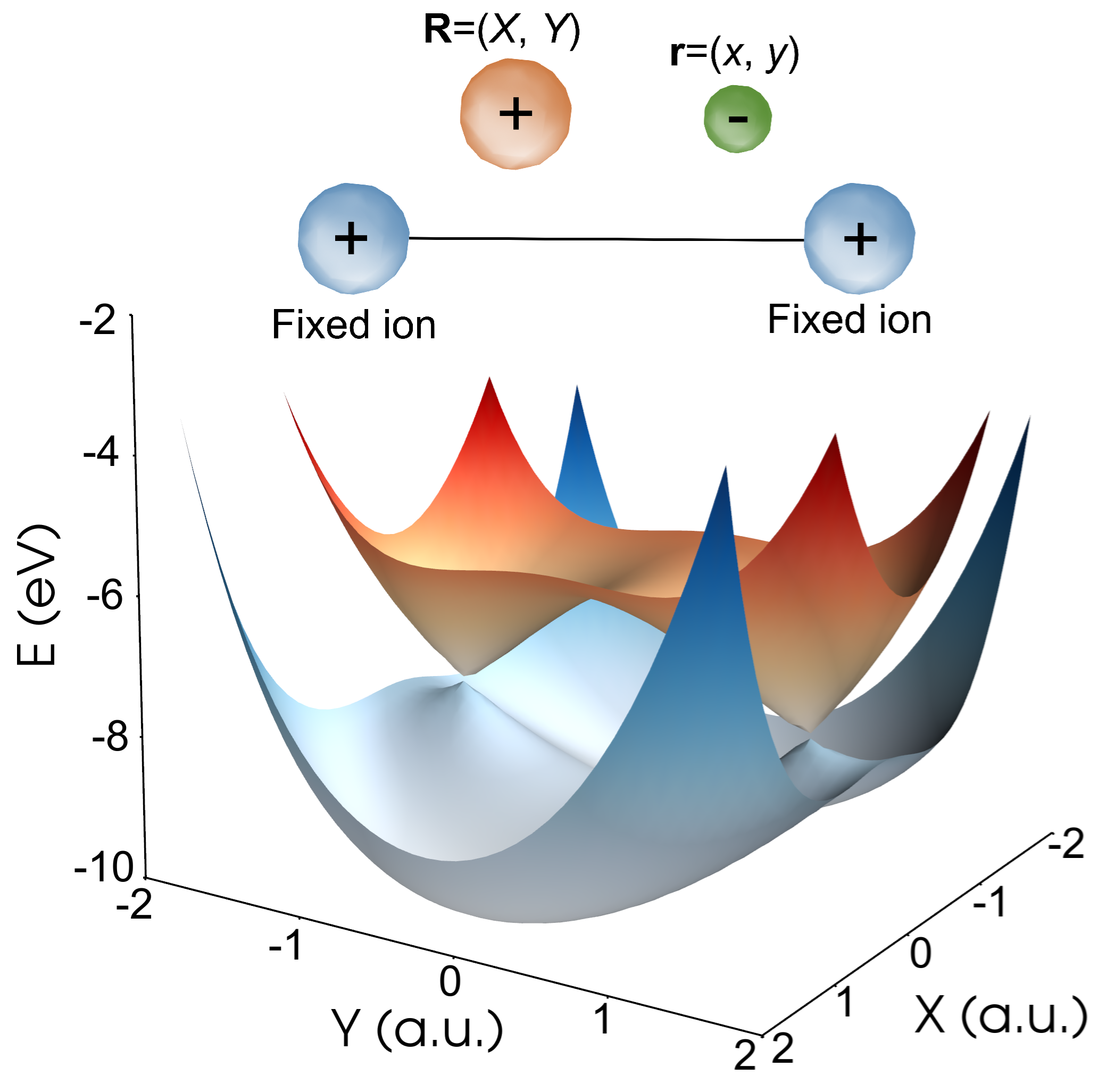}
	\caption{ The schematic and the adiabatic potential energy surfaces of the two excited states for 2D Shin-Metiu model. The parameters are chosen as (atomic units) $L = 4\sqrt{3} \text{ Bohr}, R_0=3.5 $ Bohr, $a=0.5, b=10$. }
	\label{ShinMetiu_PES}
\end{figure}

To illustrate the utility of the approximate electronic overlap matrix, we compare the conical intersection dynamics using the approximate overlap with the exact results for the two-dimensional (2D) Shin-Metiu model \cite{min2014,shin1995}. This model, widely used for modeling photoinduced proton-coupled electron transfer, consists of a proton $\bf R =(X,Y)$ and an electron $\bf r=(x,y)$ moving within a 2D plane under the soft Coulomb potential of two fixed ions at $(\pm L/2, 0)$, schematically depicted in \cref{ShinMetiu_PES}.

The full Hamiltonian of the model is given by \cite{zhu2024}

\begin{equation}
	\hat{H}(\bf{r}, \bf{R})=\frac{\bf{\hat{p}}^2}{2}+\frac{\bf{\hat{P}}^2}{2M}+\hat{V}(\bf{r}, \bf{R})
\end{equation}
where $ \bf{\hat{p}} = -i\grad_{\bf r}$ and $\bf{\hat{P}} = -i \grad_{\bf R}$ are, respectively, the momenta operators of the electron and the proton, $M$ is the proton mass. The potential energy  $\hat{V}(\bf{r}; \bf{R}) = \hat{V}_\text{en} + \hat{V}_\text{nn}+ \del{{R/R_0}}^4
$
consists of the electron-nuclear soft Coulomb interaction  $  V_\text{en}(\bf{r}, \bf{R})=-1/\sqrt{a+(\bf{r}-\bf{R})^2} - \sum_{i=1,2}  1/\sqrt{a+(\bf{r}-\bf{R}_i)^2}  $,  and the nucleus-nucleus repulsion
$  V_\text{nn}(\bf{R}) = \sum_{i=1,2} 1/\sqrt{b+(\bf{R}-\bf{R}_i)^2} $, 
where $\textbf{R}_i$ is the fixed ion position. The quartic potential is added to make the system bound \cite{min2014}. 

The  electronic states and energies $V_{\alpha}(\bf{R}_{\bf n})$ are obtained by solving the electronic \se\ in a direct product DVR basis set 
\be  
\del{ \frac{\bf{\hat{p}}^2}{2} + \hat{V}(\bf{r}; \bf{R}) } \phi_\alpha(\bf{r}; \bf{R})=V_\alpha(\bf{R}) \phi_\alpha(\bf{r}; \bf{R}).
  \ee

We use the sine DVR (particle-in-a-box eigenstates) with a two-dimensional uniform grid with 129 points in the range of $[-6, 6]$ Bohr for each electronic degree of freedom.

For each nuclear degree of freedom, the range is $[-3, 3]$ Bohr with 63 grid points. By going through the nuclear DVR points $\bf{R}_n$, the first and second excited electronic potential energy surfaces are constructed; the two surfaces intersect at $(0, \pm 1.2)$ Bohr (\fig{ShinMetiu_PES}). The initial wave packet is chosen as a Gaussian wave packet centered at (0, 2.25) Bohr in the second excited state.

\begin{figure}[htbp]
	\centering
	\includegraphics[width=0.48\textwidth]{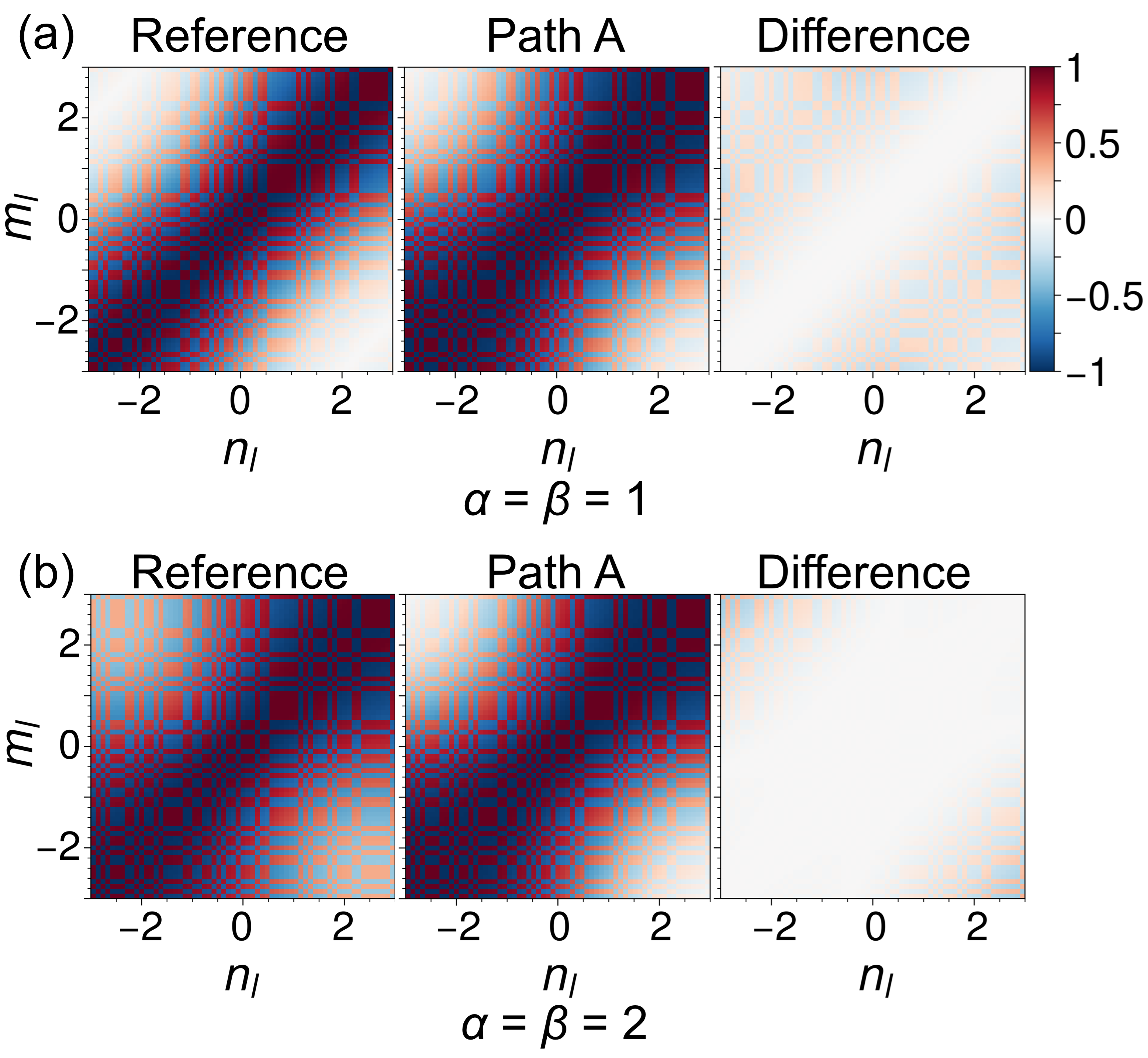}
	\caption{The exact (left) and approximate (middle) overlap matrices for (a) the first and (b) the second excited electronic states when $m_2 = n_2 = -2.91$ Bohr. The right panels denote the difference between the exact and approximate overlap matrices.}
	\label{A_matrix}
\end{figure}

\fig{A_matrix} compares the approximate intra-state overlap matrices $A^{\alpha \alpha}({ m_2=n_2=-2.91 \text{ Bohr}, \alpha=1, 2})$ to the reference matrix, obtained by exact electronic structure calculations. The random occurrence of positive and negative matrix elements is due to the randomness in the phase of adiabatic electronic wavefunctions resulting from electronic structure calculations \cite{zhu2024}. As shown, the approximate intra-state overlap matrix (middle panels) is globally phase-consistent and in good agreement with the reference one (left panels), with small differences when the geometries are far away from each other. To more clearly illustrate the differences between the approximate overlap matrix and the exact one, we have depicted them in the right panels of \fig{A_matrix}, showing that the two matrices are nearly identical for short-range overlap, with slight differences for long-range ones.
This clearly illustrates the validity of the linked product approximation. The linked product approximation only requires computation time of ~0.61 CPU hours, compared to $\sim$15.03 hours for the exact calculations.

\begin{figure}[htbp]
	\centering
	\includegraphics[width=0.48\textwidth]{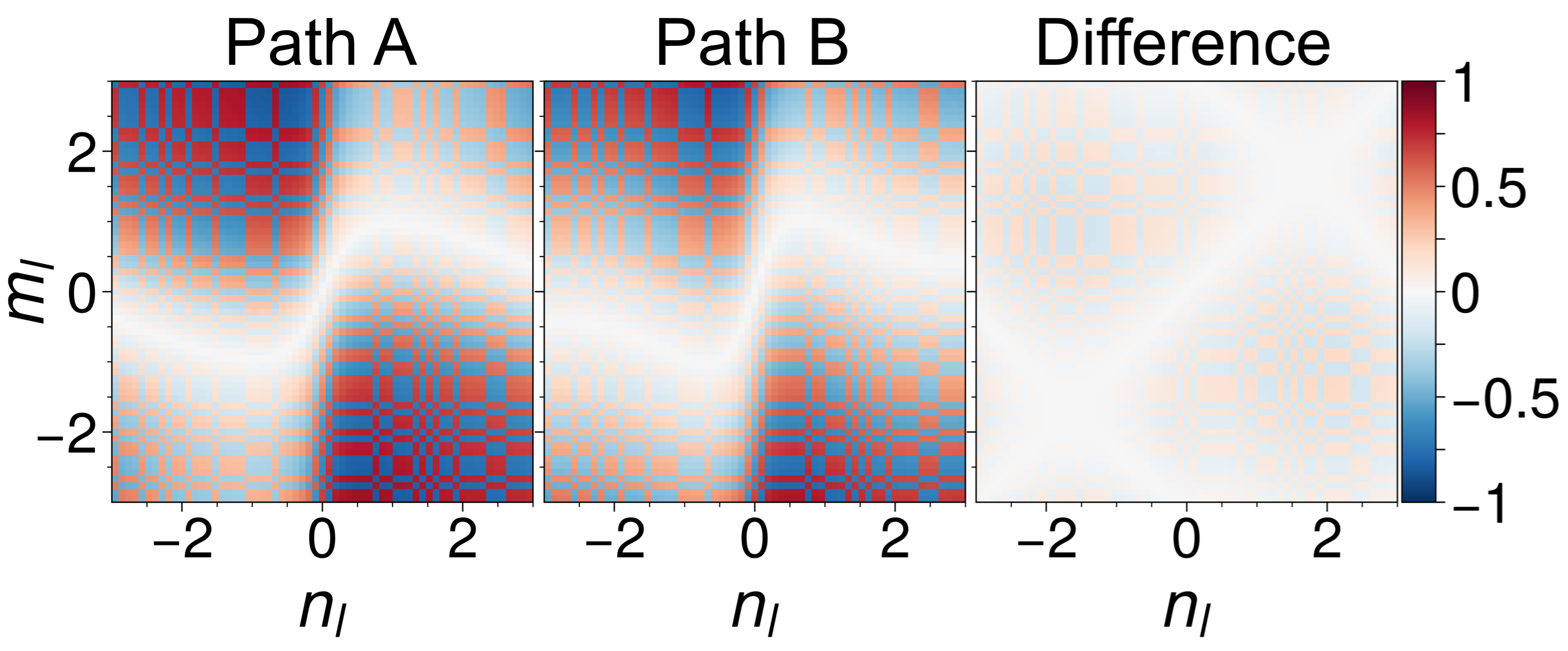}
	\caption{The approximate electronic overlap matrices of the first excited state calculated through Path A (left) and Path B (middle) when  $m_2 = -2.91 \text{ Bohr}$ and $n_2 = 0.84 \text{ Bohr}$. The right panel denotes the difference between these two approximated overlap matrices.}
	\label{difference_pahtAB}
\end{figure}

Additionally, \fig{difference_pahtAB} compares the approximate intra-state overlap matrices $A^{\alpha \alpha}$ ($m_2 = -2.91 \text{ Bohr}, n_2 = 0.84 \text{ Bohr}$ and $\alpha = 1$) of the first excited state via Path A (left) and Path B (middle), showing that the approximate overlap matrix is path-dependent. The two approximate intra-state matrices give the same phase information, but they show slight difference in the non-diagonal element. The average difference is $\sim$0.03, with the maximum difference is $\sim$0.31. These differences  appear because that different configurations involved in the two paths lead to differences in the electron projection operator $\hat{P}_{\bf n}$, making the approximate overlap matrix different. As a result, the approximate overlap matrix is path-dependent.

\begin{figure}[htbp]
	\centering
	\includegraphics[width=0.45\textwidth]{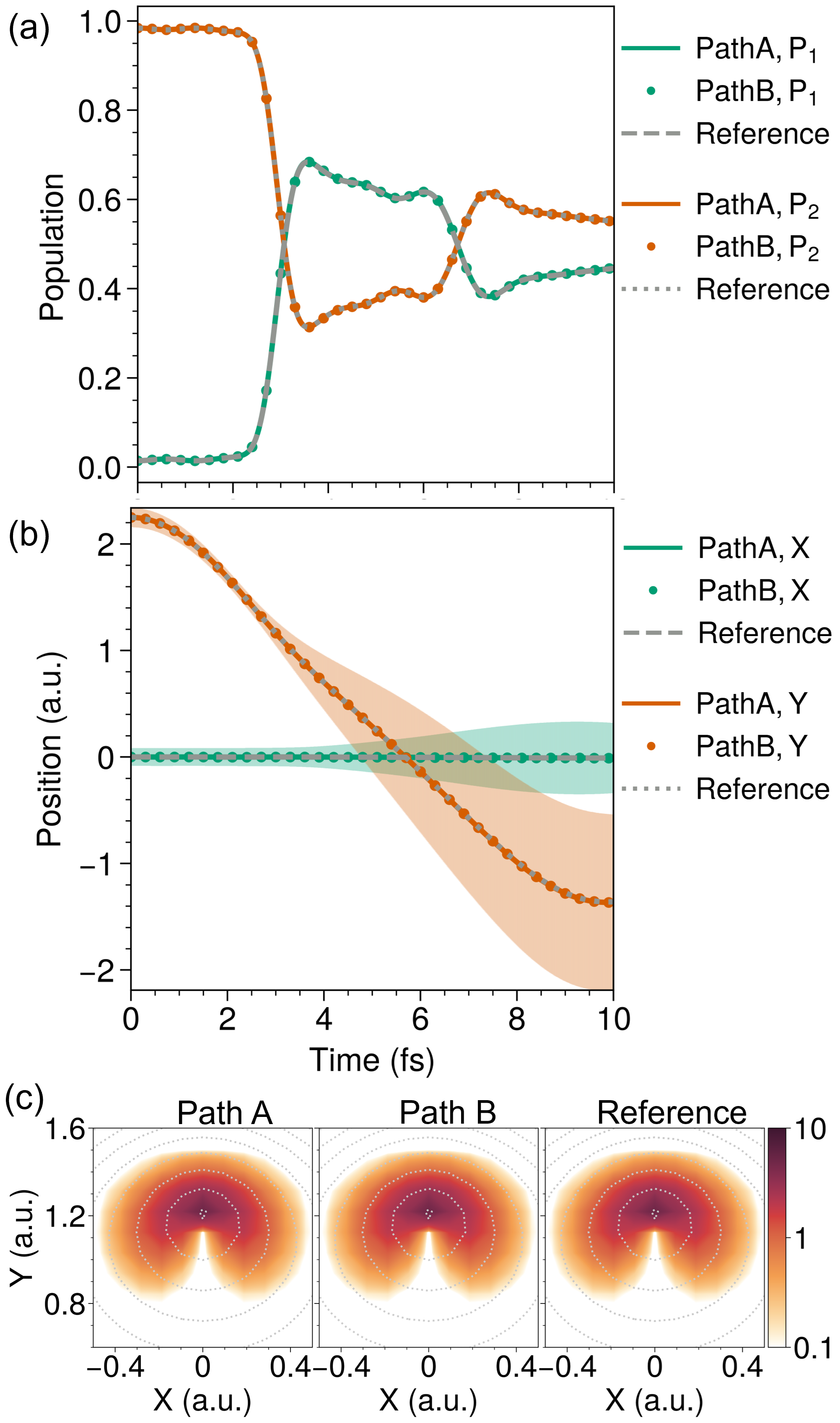}
	\caption{(a) The electronic population on two excited states and (b) the expectation value of position operator. (c) The nuclear wave packet at 3 fs obtained by the linked product approximation through Path A (left), Path B (right) and rigorous computation.}
	\label{compare_observables}
\end{figure}

\begin{figure}[htbp]
	\centering
	\includegraphics[width=0.48\textwidth]{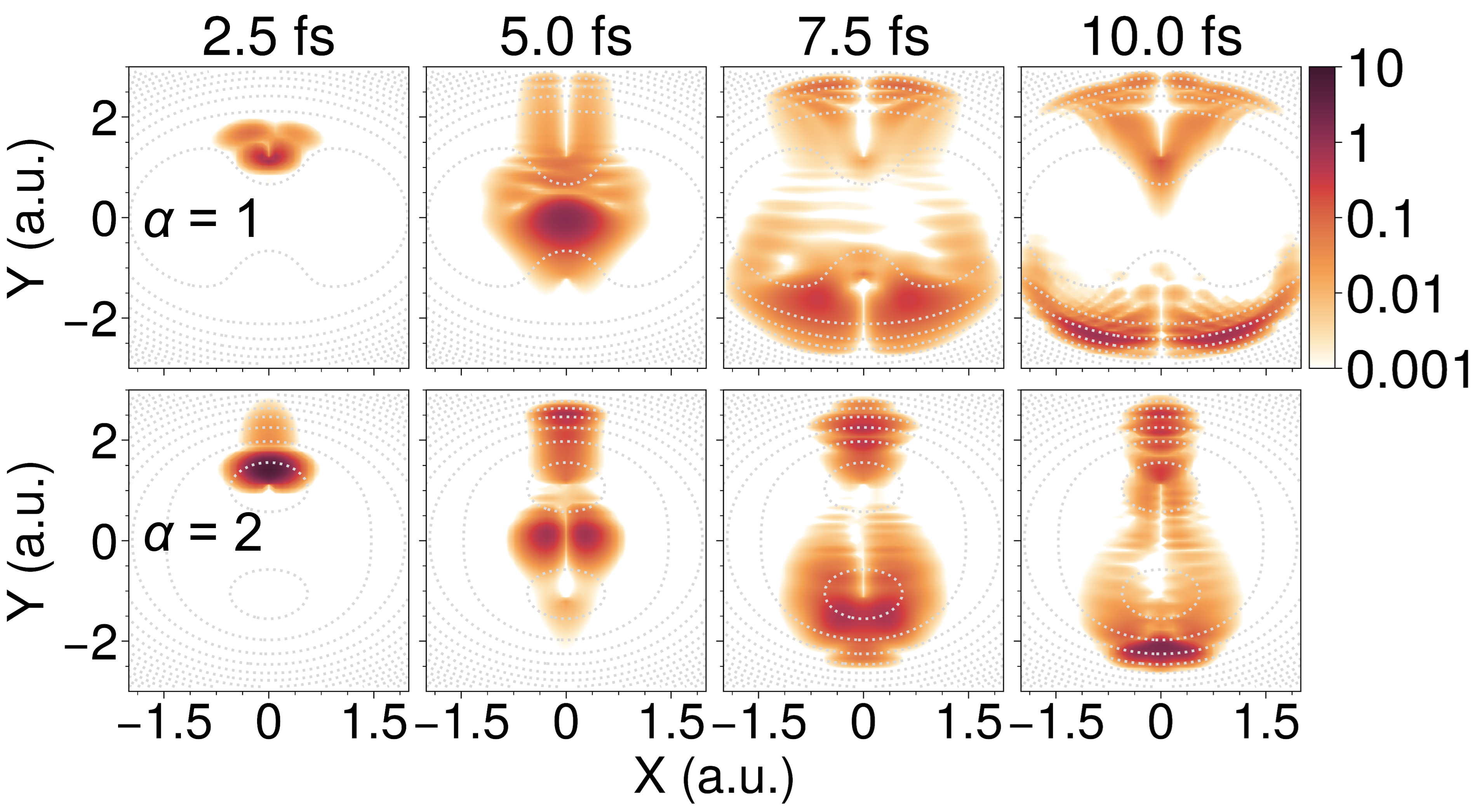}
	\caption{The nuclear wave packet distribution of the first (top, $\alpha=1$) and the second (bottom, $\alpha=2$) excited states through the conical intersection.}
	\label{wavepackets}
\end{figure}

To demonstrate that the approximate overlap matrix can yield the same quantum dynamical results as the reference overlap matrix, \fig{compare_observables} presents the electronic population dynamics, the expectation value of the position operator and the nuclear wave packet calculated using both linked product approximation and reference calculations. As plotted in \fig{compare_observables}a, the nuclear wave packet in the second excited state reaches the conical intersection at $\sim$2.5 fs, resulting in a rapid nonadiabatic transition to the lower adiabatic state. This transition process is completed within 4 fs. Subsequently, part of the nuclear wave packet goes back to the second excited states via the conical intersection.
According to the position of nuclear wave packet shown in \fig{compare_observables}b, the proton locates at the central position between the two fixed ions ($X = 0$) and moves along the minus $Y$-directions. The shaded orange and green regions indicate the variance of the position ($\sqrt{\braket{O^2}-\braket{O}^2}$), implying that the nuclear wave packet spreads out with time. The wave packet distribution within 10 fs are depicted in \fig{wavepackets}, which also shows the position of proton and its increasing variance with time.

Both the electronic population dynamics and the proton position calculated using the approximate overlap matrix perfectly match those obtained from the reference overlap matrix (see \fig{compare_observables}a and \fig{compare_observables}b). Moreover, the nuclear wave packet derived from both matrices are identical (see \fig{compare_observables}c). What'more, the wave packet distribution at 3 fs show a node at $X=0$, signifying that our approximation captures the geometric phase effect.
The consistency of quantum dynamical results suggests that, although differences exist between the approximate and reference overlap matrices in the long-range overlaps, these differences have a negligible impact on the quantum dynamics.These results further demonstrate that the linked product approximation can yield accurate outcomes while significantly reducing computation time.
Additionally, the dynamical results from the two paths are identical, suggesting that although the approximate overlap matrices differ between the paths, the quantum dynamics are path-independent.

\section{Conclusion}\label{sec:summary}



We have developed and validated the linked product approximation for efficiently calculating the electronic overlap matrix. This approach requires only the electronic structure calculations of the nearest-neighbor overlaps, while the remaining elements are obtained from their products, reducing the scaling of the computational cost from $\mc{O}(n^{2d})$ to $\mc{O}(dn^d)$.

Simulations using the two-dimensional Shin-Metiu model demonstrate that the approximate overlap matrix perfectly matches the exact values for short-range geometries. The accuracy of the approximation deteriorates as the two configurations move further apart. However, the approximate overlap matrix provides near-exact results for conical intersection dynamics, including both population dynamics and geometric phase effects. Furthermore, we showed that the path dependence of the linked product approximation is immaterial for conical intersection dynamics.

\section{Acknowledgment}
This work is supported by the National Natural Science Foundation of China (Grant No. 22473090 and 92356310).

\bibliography{dynamics, qchem, optics, cavity}

\begin{thebibliography}{22}%
\makeatletter
\providecommand \@ifxundefined [1]{%
 \@ifx{#1\undefined}
}%
\providecommand \@ifnum [1]{%
 \ifnum #1\expandafter \@firstoftwo
 \else \expandafter \@secondoftwo
 \fi
}%
\providecommand \@ifx [1]{%
 \ifx #1\expandafter \@firstoftwo
 \else \expandafter \@secondoftwo
 \fi
}%
\providecommand \natexlab [1]{#1}%
\providecommand \enquote  [1]{``#1''}%
\providecommand \bibnamefont  [1]{#1}%
\providecommand \bibfnamefont [1]{#1}%
\providecommand \citenamefont [1]{#1}%
\providecommand \href@noop [0]{\@secondoftwo}%
\providecommand \href [0]{\begingroup \@sanitize@url \@href}%
\providecommand \@href[1]{\@@startlink{#1}\@@href}%
\providecommand \@@href[1]{\endgroup#1\@@endlink}%
\providecommand \@sanitize@url [0]{\catcode `\\12\catcode `\$12\catcode
  `\&12\catcode `\#12\catcode `\^12\catcode `\_12\catcode `\%12\relax}%
\providecommand \@@startlink[1]{}%
\providecommand \@@endlink[0]{}%
\providecommand \url  [0]{\begingroup\@sanitize@url \@url }%
\providecommand \@url [1]{\endgroup\@href {#1}{\urlprefix }}%
\providecommand \urlprefix  [0]{URL }%
\providecommand \Eprint [0]{\href }%
\providecommand \doibase [0]{https://doi.org/}%
\providecommand \selectlanguage [0]{\@gobble}%
\providecommand \bibinfo  [0]{\@secondoftwo}%
\providecommand \bibfield  [0]{\@secondoftwo}%
\providecommand \translation [1]{[#1]}%
\providecommand \BibitemOpen [0]{}%
\providecommand \bibitemStop [0]{}%
\providecommand \bibitemNoStop [0]{.\EOS\space}%
\providecommand \EOS [0]{\spacefactor3000\relax}%
\providecommand \BibitemShut  [1]{\csname bibitem#1\endcsname}%
\let\auto@bib@innerbib\@empty
\bibitem [{\citenamefont {Domcke}\ \emph {et~al.}(2011)\citenamefont {Domcke},
  \citenamefont {Yarkony},\ and\ \citenamefont {K{\"o}ppel}}]{domcke2011}%
  \BibitemOpen
  \bibfield  {author} {\bibinfo {author} {\bibfnamefont {W.}~\bibnamefont
  {Domcke}}, \bibinfo {author} {\bibfnamefont {D.~R.}\ \bibnamefont
  {Yarkony}},\ and\ \bibinfo {author} {\bibfnamefont {H.}~\bibnamefont
  {K{\"o}ppel}},\ }\href@noop {} {\emph {\bibinfo {title} {Conical
  {{Intersections}}: {{Theory}}, {{Computation}} and {{Experiment}}}}}\
  (\bibinfo  {publisher} {World Scientific},\ \bibinfo {year}
  {2011})\BibitemShut {NoStop}%
\bibitem [{\citenamefont {Larson}\ \emph {et~al.}(2020)\citenamefont {Larson},
  \citenamefont {Sj{\"o}qvist},\ and\ \citenamefont {{\"O}hberg}}]{larson2020}%
  \BibitemOpen
  \bibfield  {author} {\bibinfo {author} {\bibfnamefont {J.}~\bibnamefont
  {Larson}}, \bibinfo {author} {\bibfnamefont {E.}~\bibnamefont
  {Sj{\"o}qvist}},\ and\ \bibinfo {author} {\bibfnamefont {P.}~\bibnamefont
  {{\"O}hberg}},\ }\href {https://doi.org/10.1007/978-3-030-34882-3} {\emph
  {\bibinfo {title} {Conical {{Intersections}} in {{Physics}}: {{An
  Introduction}} to {{Synthetic Gauge Theories}}}}},\ \bibinfo {series}
  {Lecture {{Notes}} in {{Physics}}}, Vol.\ \bibinfo {volume} {965}\ (\bibinfo
  {publisher} {Springer International Publishing},\ \bibinfo {address} {Cham},\
  \bibinfo {year} {2020})\BibitemShut {NoStop}%
\bibitem [{\citenamefont {W{\"o}rner}\ \emph {et~al.}(2011)\citenamefont
  {W{\"o}rner}, \citenamefont {Bertrand}, \citenamefont {Fabre}, \citenamefont
  {Higuet}, \citenamefont {Ruf}, \citenamefont {Dubrouil}, \citenamefont
  {Patchkovskii}, \citenamefont {Spanner}, \citenamefont {Mairesse},
  \citenamefont {Blanchet}, \citenamefont {M{\'e}vel}, \citenamefont
  {Constant}, \citenamefont {Corkum},\ and\ \citenamefont
  {Villeneuve}}]{worner2011}%
  \BibitemOpen
  \bibfield  {author} {\bibinfo {author} {\bibfnamefont {H.~J.}\ \bibnamefont
  {W{\"o}rner}}, \bibinfo {author} {\bibfnamefont {J.~B.}\ \bibnamefont
  {Bertrand}}, \bibinfo {author} {\bibfnamefont {B.}~\bibnamefont {Fabre}},
  \bibinfo {author} {\bibfnamefont {J.}~\bibnamefont {Higuet}}, \bibinfo
  {author} {\bibfnamefont {H.}~\bibnamefont {Ruf}}, \bibinfo {author}
  {\bibfnamefont {A.}~\bibnamefont {Dubrouil}}, \bibinfo {author}
  {\bibfnamefont {S.}~\bibnamefont {Patchkovskii}}, \bibinfo {author}
  {\bibfnamefont {M.}~\bibnamefont {Spanner}}, \bibinfo {author} {\bibfnamefont
  {Y.}~\bibnamefont {Mairesse}}, \bibinfo {author} {\bibfnamefont
  {V.}~\bibnamefont {Blanchet}}, \bibinfo {author} {\bibfnamefont
  {E.}~\bibnamefont {M{\'e}vel}}, \bibinfo {author} {\bibfnamefont
  {E.}~\bibnamefont {Constant}}, \bibinfo {author} {\bibfnamefont {P.~B.}\
  \bibnamefont {Corkum}},\ and\ \bibinfo {author} {\bibfnamefont {D.~M.}\
  \bibnamefont {Villeneuve}},\ }\bibfield  {title} {\bibinfo {title} {Conical
  {{Intersection Dynamics}} in {{NO2 Probed}} by {{Homodyne High-Harmonic
  Spectroscopy}}},\ }\href {https://doi.org/10.1126/science.1208664} {\bibfield
   {journal} {\bibinfo  {journal} {Science}\ }\textbf {\bibinfo {volume}
  {334}},\ \bibinfo {pages} {208} (\bibinfo {year} {2011})}\BibitemShut
  {NoStop}%
\bibitem [{\citenamefont {Kowalewski}\ and\ \citenamefont
  {Mukamel}(2017)}]{kowalewski2017}%
  \BibitemOpen
  \bibfield  {author} {\bibinfo {author} {\bibfnamefont {M.}~\bibnamefont
  {Kowalewski}}\ and\ \bibinfo {author} {\bibfnamefont {S.}~\bibnamefont
  {Mukamel}},\ }\bibfield  {title} {\bibinfo {title} {Manipulating molecules
  with quantum light},\ }\href {https://doi.org/10.1073/pnas.1702160114}
  {\bibfield  {journal} {\bibinfo  {journal} {Proc. Natl. Acad. Sci.}\ }\textbf
  {\bibinfo {volume} {114}},\ \bibinfo {pages} {3278} (\bibinfo {year}
  {2017})}\BibitemShut {NoStop}%
\bibitem [{\citenamefont {Gu}\ and\ \citenamefont
  {Mukamel}(2020{\natexlab{a}})}]{gu2020b}%
  \BibitemOpen
  \bibfield  {author} {\bibinfo {author} {\bibfnamefont {B.}~\bibnamefont
  {Gu}}\ and\ \bibinfo {author} {\bibfnamefont {S.}~\bibnamefont {Mukamel}},\
  }\bibfield  {title} {\bibinfo {title} {Cooperative {{Conical Intersection
  Dynamics}} of {{Two Pyrazine Molecules}} in an {{Optical Cavity}}},\ }\href
  {https://doi.org/10.1021/acs.jpclett.0c00381} {\bibfield  {journal} {\bibinfo
   {journal} {J. Phys. Chem. Lett.}\ }\textbf {\bibinfo {volume} {11}},\
  \bibinfo {pages} {5555} (\bibinfo {year} {2020}{\natexlab{a}})}\BibitemShut
  {NoStop}%
\bibitem [{\citenamefont {Gu}\ and\ \citenamefont
  {Mukamel}(2020{\natexlab{b}})}]{gu2020c}%
  \BibitemOpen
  \bibfield  {author} {\bibinfo {author} {\bibfnamefont {B.}~\bibnamefont
  {Gu}}\ and\ \bibinfo {author} {\bibfnamefont {S.}~\bibnamefont {Mukamel}},\
  }\bibfield  {title} {\bibinfo {title} {Manipulating nonadiabatic conical
  intersection dynamics by optical cavities},\ }\href
  {https://doi.org/10.1039/C9SC04992D} {\bibfield  {journal} {\bibinfo
  {journal} {Chem. Sci.}\ }\textbf {\bibinfo {volume} {11}},\ \bibinfo {pages}
  {1290} (\bibinfo {year} {2020}{\natexlab{b}})}\BibitemShut {NoStop}%
\bibitem [{\citenamefont {Yang}\ \emph {et~al.}(2020)\citenamefont {Yang},
  \citenamefont {Khosravi},\ and\ \citenamefont {Jacob}}]{yang2020b}%
  \BibitemOpen
  \bibfield  {author} {\bibinfo {author} {\bibfnamefont {L.-P.}\ \bibnamefont
  {Yang}}, \bibinfo {author} {\bibfnamefont {F.}~\bibnamefont {Khosravi}},\
  and\ \bibinfo {author} {\bibfnamefont {Z.}~\bibnamefont {Jacob}},\ }\bibfield
   {title} {\bibinfo {title} {Quantum spin operator of the photon},\
  }\href@noop {} {\bibfield  {journal} {\bibinfo  {journal} {arXiv:2004.03771
  [cond-mat, physics:hep-th, physics:physics, physics:quant-ph]}\ } (\bibinfo
  {year} {2020})},\ \Eprint {https://arxiv.org/abs/2004.03771}
  {arXiv:2004.03771 [cond-mat, physics:hep-th, physics:physics,
  physics:quant-ph]} \BibitemShut {NoStop}%
\bibitem [{\citenamefont {Xie}\ \emph {et~al.}(2016)\citenamefont {Xie},
  \citenamefont {Ma}, \citenamefont {Zhu}, \citenamefont {Yarkony},
  \citenamefont {Xie},\ and\ \citenamefont {Guo}}]{xie2016}%
  \BibitemOpen
  \bibfield  {author} {\bibinfo {author} {\bibfnamefont {C.}~\bibnamefont
  {Xie}}, \bibinfo {author} {\bibfnamefont {J.}~\bibnamefont {Ma}}, \bibinfo
  {author} {\bibfnamefont {X.}~\bibnamefont {Zhu}}, \bibinfo {author}
  {\bibfnamefont {D.~R.}\ \bibnamefont {Yarkony}}, \bibinfo {author}
  {\bibfnamefont {D.}~\bibnamefont {Xie}},\ and\ \bibinfo {author}
  {\bibfnamefont {H.}~\bibnamefont {Guo}},\ }\bibfield  {title} {\bibinfo
  {title} {Nonadiabatic {{Tunneling}} in {{Photodissociation}} of {{Phenol}}},\
  }\href {https://doi.org/10.1021/jacs.6b03288} {\bibfield  {journal} {\bibinfo
   {journal} {J. Am. Chem. Soc.}\ }\textbf {\bibinfo {volume} {138}},\ \bibinfo
  {pages} {7828} (\bibinfo {year} {2016})}\BibitemShut {NoStop}%
\bibitem [{\citenamefont {Han}\ \emph {et~al.}(2023)\citenamefont {Han},
  \citenamefont {Xie}, \citenamefont {Hu}, \citenamefont {Yarkony},
  \citenamefont {Guo},\ and\ \citenamefont {Xie}}]{han2023}%
  \BibitemOpen
  \bibfield  {author} {\bibinfo {author} {\bibfnamefont {S.}~\bibnamefont
  {Han}}, \bibinfo {author} {\bibfnamefont {C.}~\bibnamefont {Xie}}, \bibinfo
  {author} {\bibfnamefont {X.}~\bibnamefont {Hu}}, \bibinfo {author}
  {\bibfnamefont {D.~R.}\ \bibnamefont {Yarkony}}, \bibinfo {author}
  {\bibfnamefont {H.}~\bibnamefont {Guo}},\ and\ \bibinfo {author}
  {\bibfnamefont {D.}~\bibnamefont {Xie}},\ }\bibfield  {title} {\bibinfo
  {title} {Quantum {{Dynamics}} of {{Photodissociation}}: {{Recent Advances}}
  and {{Challenges}}},\ }\href {https://doi.org/10.1021/acs.jpclett.3c02735}
  {\bibfield  {journal} {\bibinfo  {journal} {J. Phys. Chem. Lett.}\ }\textbf
  {\bibinfo {volume} {14}},\ \bibinfo {pages} {10517} (\bibinfo {year}
  {2023})}\BibitemShut {NoStop}%
\bibitem [{\citenamefont {Zhu}\ and\ \citenamefont {Gu}(2024)}]{zhu2024}%
  \BibitemOpen
  \bibfield  {author} {\bibinfo {author} {\bibfnamefont {X.}~\bibnamefont
  {Zhu}}\ and\ \bibinfo {author} {\bibfnamefont {B.}~\bibnamefont {Gu}},\
  }\bibfield  {title} {\bibinfo {title} {Making {{Peace}} with {{Random
  Phases}}: {{Ab Initio Conical Intersection Quantum Dynamics}} in {{Random
  Gauges}}},\ }\href {https://doi.org/10.1021/acs.jpclett.4c01688} {\bibfield
  {journal} {\bibinfo  {journal} {J. Phys. Chem. Lett.}\ }\textbf {\bibinfo
  {volume} {15}},\ \bibinfo {pages} {8487} (\bibinfo {year}
  {2024})}\BibitemShut {NoStop}%
\bibitem [{\citenamefont {Aleotti}\ \emph {et~al.}(2021)\citenamefont
  {Aleotti}, \citenamefont {Aranda}, \citenamefont {Yaghoubi~Jouybari},
  \citenamefont {Garavelli}, \citenamefont {Nenov},\ and\ \citenamefont
  {Santoro}}]{aleotti2021b}%
  \BibitemOpen
  \bibfield  {author} {\bibinfo {author} {\bibfnamefont {F.}~\bibnamefont
  {Aleotti}}, \bibinfo {author} {\bibfnamefont {D.}~\bibnamefont {Aranda}},
  \bibinfo {author} {\bibfnamefont {M.}~\bibnamefont {Yaghoubi~Jouybari}},
  \bibinfo {author} {\bibfnamefont {M.}~\bibnamefont {Garavelli}}, \bibinfo
  {author} {\bibfnamefont {A.}~\bibnamefont {Nenov}},\ and\ \bibinfo {author}
  {\bibfnamefont {F.}~\bibnamefont {Santoro}},\ }\bibfield  {title} {\bibinfo
  {title} {Parameterization of a linear vibronic coupling model with
  multiconfigurational electronic structure methods to study the quantum
  dynamics of photoexcited pyrene},\ }\href {https://doi.org/10.1063/5.0044693}
  {\bibfield  {journal} {\bibinfo  {journal} {J. Chem. Phys.}\ }\textbf
  {\bibinfo {volume} {154}},\ \bibinfo {pages} {104106} (\bibinfo {year}
  {2021})}\BibitemShut {NoStop}%
\bibitem [{\citenamefont {Shu}\ \emph {et~al.}(2022)\citenamefont {Shu},
  \citenamefont {Varga}, \citenamefont {Kanchanakungwankul}, \citenamefont
  {Zhang},\ and\ \citenamefont {Truhlar}}]{shu2022}%
  \BibitemOpen
  \bibfield  {author} {\bibinfo {author} {\bibfnamefont {Y.}~\bibnamefont
  {Shu}}, \bibinfo {author} {\bibfnamefont {Z.}~\bibnamefont {Varga}}, \bibinfo
  {author} {\bibfnamefont {S.}~\bibnamefont {Kanchanakungwankul}}, \bibinfo
  {author} {\bibfnamefont {L.}~\bibnamefont {Zhang}},\ and\ \bibinfo {author}
  {\bibfnamefont {D.~G.}\ \bibnamefont {Truhlar}},\ }\bibfield  {title}
  {\bibinfo {title} {Diabatic {{States}} of {{Molecules}}},\ }\href
  {https://doi.org/10.1021/acs.jpca.1c10583} {\bibfield  {journal} {\bibinfo
  {journal} {J. Phys. Chem. A}\ }\textbf {\bibinfo {volume} {126}},\ \bibinfo
  {pages} {992} (\bibinfo {year} {2022})}\BibitemShut {NoStop}%
\bibitem [{\citenamefont {Gu}(2023)}]{gu2023b}%
  \BibitemOpen
  \bibfield  {author} {\bibinfo {author} {\bibfnamefont {B.}~\bibnamefont
  {Gu}},\ }\bibfield  {title} {\bibinfo {title} {A {{Discrete-Variable Local
  Diabatic Representation}} of {{Conical Intersection Dynamics}}},\ }\href
  {https://doi.org/10.1021/acs.jctc.3c00560} {\bibfield  {journal} {\bibinfo
  {journal} {J. Chem. Theory Comput.}\ }\textbf {\bibinfo {volume} {19}},\
  \bibinfo {pages} {6557} (\bibinfo {year} {2023})}\BibitemShut {NoStop}%
\bibitem [{\citenamefont {Gu}(2024)}]{gu2024a}%
  \BibitemOpen
  \bibfield  {author} {\bibinfo {author} {\bibfnamefont {B.}~\bibnamefont
  {Gu}},\ }\bibfield  {title} {\bibinfo {title} {Nonadiabatic {{Conical
  Intersection Dynamics}} in the {{Local Diabatic Representation}} with
  {{Strang Splitting}} and {{Fourier Basis}}},\ }\href
  {https://doi.org/10.1021/acs.jctc.3c01317} {\bibfield  {journal} {\bibinfo
  {journal} {J. Chem. Theory Comput.}\ }\textbf {\bibinfo {volume} {20}},\
  \bibinfo {pages} {2711} (\bibinfo {year} {2024})}\BibitemShut {NoStop}%
\bibitem [{\citenamefont {Shin}\ and\ \citenamefont {Metiu}(1995)}]{shin1995}%
  \BibitemOpen
  \bibfield  {author} {\bibinfo {author} {\bibfnamefont {S.}~\bibnamefont
  {Shin}}\ and\ \bibinfo {author} {\bibfnamefont {H.}~\bibnamefont {Metiu}},\
  }\bibfield  {title} {\bibinfo {title} {Nonadiabatic effects on the charge
  transfer rate constant: {{A}} numerical study of a simple model system},\
  }\href {https://doi.org/10.1063/1.468795} {\bibfield  {journal} {\bibinfo
  {journal} {The Journal of Chemical Physics}\ }\textbf {\bibinfo {volume}
  {102}},\ \bibinfo {pages} {9285} (\bibinfo {year} {1995})}\BibitemShut
  {NoStop}%
\bibitem [{\citenamefont {Shin}\ and\ \citenamefont {Cho}(2000)}]{shin2000}%
  \BibitemOpen
  \bibfield  {author} {\bibinfo {author} {\bibfnamefont {S.}~\bibnamefont
  {Shin}}\ and\ \bibinfo {author} {\bibfnamefont {S.-I.}\ \bibnamefont {Cho}},\
  }\bibfield  {title} {\bibinfo {title} {Quantum dynamics of model
  proton-coupled electron transfer reactions},\ }\href
  {https://doi.org/10.1016/S0301-0104(00)00211-1} {\bibfield  {journal}
  {\bibinfo  {journal} {Chem. Phys.}\ }\textbf {\bibinfo {volume} {259}},\
  \bibinfo {pages} {27} (\bibinfo {year} {2000})}\BibitemShut {NoStop}%
\bibitem [{\citenamefont {Light}\ and\ \citenamefont
  {Carrington~Jr.}(2000)}]{light2000}%
  \BibitemOpen
  \bibfield  {author} {\bibinfo {author} {\bibfnamefont {J.~C.}\ \bibnamefont
  {Light}}\ and\ \bibinfo {author} {\bibfnamefont {T.}~\bibnamefont
  {Carrington~Jr.}},\ }\bibfield  {title} {\bibinfo {title}
  {Discrete-{{Variable Representations}} and their {{Utilization}}},\ }in\
  \href {https://doi.org/10.1002/9780470141731.ch4} {\emph {\bibinfo
  {booktitle} {Advances in {{Chemical Physics}}}}}\ (\bibinfo  {publisher}
  {John Wiley \& Sons, Ltd},\ \bibinfo {year} {2000})\ pp.\ \bibinfo {pages}
  {263--310}\BibitemShut {NoStop}%
\bibitem [{\citenamefont {Littlejohn}\ \emph {et~al.}(2002)\citenamefont
  {Littlejohn}, \citenamefont {Cargo}, \citenamefont {Carrington},
  \citenamefont {Mitchell},\ and\ \citenamefont {Poirier}}]{littlejohn2002}%
  \BibitemOpen
  \bibfield  {author} {\bibinfo {author} {\bibfnamefont {R.~G.}\ \bibnamefont
  {Littlejohn}}, \bibinfo {author} {\bibfnamefont {M.}~\bibnamefont {Cargo}},
  \bibinfo {author} {\bibfnamefont {T.}~\bibnamefont {Carrington}}, \bibinfo
  {author} {\bibfnamefont {K.~A.}\ \bibnamefont {Mitchell}},\ and\ \bibinfo
  {author} {\bibfnamefont {B.}~\bibnamefont {Poirier}},\ }\bibfield  {title}
  {\bibinfo {title} {A general framework for discrete variable representation
  basis sets},\ }\href {https://doi.org/10.1063/1.1473811} {\bibfield
  {journal} {\bibinfo  {journal} {J. Chem. Phys.}\ }\textbf {\bibinfo {volume}
  {116}},\ \bibinfo {pages} {8691} (\bibinfo {year} {2002})}\BibitemShut
  {NoStop}%
\bibitem [{\citenamefont {Gu}\ \emph {et~al.}(2024)\citenamefont {Gu},
  \citenamefont {Yong}, \citenamefont {Gu},\ and\ \citenamefont
  {Mukamel}}]{gu2024}%
  \BibitemOpen
  \bibfield  {author} {\bibinfo {author} {\bibfnamefont {Y.}~\bibnamefont
  {Gu}}, \bibinfo {author} {\bibfnamefont {H.}~\bibnamefont {Yong}}, \bibinfo
  {author} {\bibfnamefont {B.}~\bibnamefont {Gu}},\ and\ \bibinfo {author}
  {\bibfnamefont {S.}~\bibnamefont {Mukamel}},\ }\bibfield  {title} {\bibinfo
  {title} {Chemical bond reorganization in intramolecular proton transfer
  revealed by ultrafast {{X-ray}} photoelectron spectroscopy},\ }\href
  {https://doi.org/10.1073/pnas.2321343121} {\bibfield  {journal} {\bibinfo
  {journal} {Proceedings of the National Academy of Sciences}\ }\textbf
  {\bibinfo {volume} {121}},\ \bibinfo {pages} {e2321343121} (\bibinfo {year}
  {2024})}\BibitemShut {NoStop}%
\bibitem [{\citenamefont {Lu}\ and\ \citenamefont {Gao}(2022)}]{lu2022}%
  \BibitemOpen
  \bibfield  {author} {\bibinfo {author} {\bibfnamefont {Y.}~\bibnamefont
  {Lu}}\ and\ \bibinfo {author} {\bibfnamefont {J.}~\bibnamefont {Gao}},\
  }\bibfield  {title} {\bibinfo {title} {Multistate {{Density Functional
  Theory}} of {{Excited States}}},\ }\href
  {https://doi.org/10.1021/acs.jpclett.2c02088} {\bibfield  {journal} {\bibinfo
   {journal} {J. Phys. Chem. Lett.}\ }\textbf {\bibinfo {volume} {13}},\
  \bibinfo {pages} {7762} (\bibinfo {year} {2022})}\BibitemShut {NoStop}%
\bibitem [{\citenamefont {Yang}\ \emph {et~al.}(2016)\citenamefont {Yang},
  \citenamefont {Shen}, \citenamefont {Zhang},\ and\ \citenamefont
  {Yang}}]{yang2016}%
  \BibitemOpen
  \bibfield  {author} {\bibinfo {author} {\bibfnamefont {Y.}~\bibnamefont
  {Yang}}, \bibinfo {author} {\bibfnamefont {L.}~\bibnamefont {Shen}}, \bibinfo
  {author} {\bibfnamefont {D.}~\bibnamefont {Zhang}},\ and\ \bibinfo {author}
  {\bibfnamefont {W.}~\bibnamefont {Yang}},\ }\bibfield  {title} {\bibinfo
  {title} {Conical {{Intersections}} from {{Particle}}--{{Particle Random
  Phase}} and {{Tamm}}--{{Dancoff Approximations}}},\ }\href
  {https://doi.org/10.1021/acs.jpclett.6b00936} {\bibfield  {journal} {\bibinfo
   {journal} {J. Phys. Chem. Lett.}\ }\textbf {\bibinfo {volume} {7}},\
  \bibinfo {pages} {2407} (\bibinfo {year} {2016})}\BibitemShut {NoStop}%
\bibitem [{\citenamefont {Min}\ \emph {et~al.}(2014)\citenamefont {Min},
  \citenamefont {Abedi}, \citenamefont {Kim},\ and\ \citenamefont
  {Gross}}]{min2014}%
  \BibitemOpen
  \bibfield  {author} {\bibinfo {author} {\bibfnamefont {S.~K.}\ \bibnamefont
  {Min}}, \bibinfo {author} {\bibfnamefont {A.}~\bibnamefont {Abedi}}, \bibinfo
  {author} {\bibfnamefont {K.~S.}\ \bibnamefont {Kim}},\ and\ \bibinfo {author}
  {\bibfnamefont {E.~K.~U.}\ \bibnamefont {Gross}},\ }\bibfield  {title}
  {\bibinfo {title} {Is the {{Molecular Berry Phase}} an {{Artifact}} of the
  {{Born-Oppenheimer Approximation}}?},\ }\href
  {https://doi.org/10.1103/PhysRevLett.113.263004} {\bibfield  {journal}
  {\bibinfo  {journal} {Phys. Rev. Lett.}\ }\textbf {\bibinfo {volume} {113}},\
  \bibinfo {pages} {263004} (\bibinfo {year} {2014})}\BibitemShut {NoStop}%
\end{thebibliography}%

\appendix

\end{document}